\documentstyle[sprocl]{article}

\def\Journal#1#2#3#4{{#1}{\bf #2}, #3 #4}

\def\NPB{{\em Nucl. Phys.} {\bf B}}
\def\PLB{{\em Phys. Lett.}  {\bf B}}
\def\PRL{\em Phys. Rev. Lett. }
\def\PRD{{\em Phys. Rev.} {\bf D}}

\def\be{\begin{equation}}
\def\ee{\end{equation}}
\def\bea{\begin{eqnarray}}
\def\eea{\end{eqnarray}}
\def\Z{{\bf Z}}

\begin{document}

\title{COSMOLOGY OF THE NEXT-TO-MINIMAL SUPERSYMMETRIC
STANDARD MODEL}

\author{ S.A. ABEL }

\address{Rutherford Appleton Laboratory, Chilton, Didcot, \\
OX11 0QX, U.K.}

\author{ S. SARKAR, P.L. WHITE\footnote{Talk presented by
P.L. White at Valencia 95} }

\address{Theoretical Physics, University of Oxford, 1 Keble Road,\\
Oxford OX1 3NP, U.K.}


\maketitle\abstracts{
We discuss the domain wall problem in the Next-to-Minimal
Supersymmetric Standard Model, with particular attention to the usual
solution of explicit breaking of the discrete symmetry by
non-renormalisable operators. This ``solution'' leads to a
contradiction between the requirements of cosmology and those of
avoiding the destabilisation of the hierarchy.
}

\section{Introduction}
One of the most popular extensions of the Minimal Supersymmetric
Standard Model (MSSM) is that of the Next-to-Minimal Supersymmetric
Standard Model (NMSSM)\cite{NMSSM} where the usual Higgs sector of two
doublets is supplemented by a gauge singlet superfield coupling only
through the Higgs sector superpotential. Apart from the important
phenomenological consequences, this allows the elimination of the $\mu$
term from the MSSM superpotential by invoking a $\Z_3$ symmetry. The
$\mu$ term presents a problem because one might expect it to be of
order the Planck mass or at least the GUT scale, and yet reasonable
phenomenology forces it to be of order the supersymmetry breaking scale
$m_{3/2}$.\cite{muprob}

The purpose of this talk is to briefly review one of the most important
cosmological implications of the NMSSM, namely that of the domain walls
which inevitably form at the weak scale when the $\Z_3$ is
spontaneously broken\cite{walls}. The requirements that these walls do
not destroy the predictions of standard cosmology forces us to
explicitly break the $\Z_3$, which can be done with non-renormalisable
operators (NROs). This in turn leads to problems with the
destabilisation of the hierarchy, and we shall discuss how the relative
sizes of the constraints affect the model. A more complete version of
this work is contained in Reference 4.

\section{The NMSSM}
In the MSSM the Higgs sector contains two doublets $H_1$ and $H_2$
coupling to down- and up-type quarks respectively, with a superpotential
containing a term  $\mu H_1H_2$. The origin of the $\mu$ term is
obscure, but we must avoid taking it much larger than the supersymmetry
breaking scale, which is typically assumed to be less than
around 1TeV.

One possible solution to the $\mu$ problem, which has the additional
advantage of giving a radically different low energy Higgs
phenomenology, is that of the NMSSM\cite{NMSSM}. Here we add a gauge
singlet superfield $N$, and replace the $\mu$ term in the superpotential
with the Higgs superpotential
\be
W_{Higgs}=\lambda NH_1H_2 - \frac{k}{3}N^3
\ee
while the usual soft breaking terms are supplemented by another mass
and two more trilinear terms instead of the bilinear term $B\mu H_1H_2$
of the MSSM. It is then possible to arrange the parameters of the model
in a natural way so that the singlet gets a vacuum expectation value
(VEV) which is of the same order as those of the other Higgs fields,
generating an effective $\mu$ parameter of form $\lambda x$ where
$x=<N>$. The terms which we have left out of the potential consistent
with gauge symmetry, namely $\mu H_1H_2$, $\mu^{\prime}N^2$, and
$\mu^{\prime\prime}N$, can all be banned by invoking a $\Z_3$ symmetry
under which all chiral superfields have the same charge so that only
trilinear terms are allowed in the superpotential. The inclusion of the
$N^3$ term is necessary both because it is consistent with $\Z_3$ and
because without it there is a Peccei-Quinn symmetry which gives a
phenomenologically unacceptable axion.

\section{Domain Walls}
\subsection{Formation and Structure}
When a discrete symmetry is spontaneously broken, it generates domain
walls\cite{walls}. This is simple to understand, since the discrete
symmetry imposes that there are multiple degenerate vacua, and there is
no way in which causally disconnected regions of the universe can
conspire to all undergo a phase transition to the same vacuum (so long
as the fields acquiring VEVs are in thermal equilibrium, which will
always be true for the NMSSM). The universe must thus evolve to a state
in which there are domains of different vacua, separated by ``domain
walls''.

On purely dimensional grounds, or by analogy with analytically solvable
toy models, we expect the domain walls to have a thickness $\delta$ and
surface energy $\sigma$ given by
\be
\delta\sim\frac{1}{\nu}
\qquad\qquad
\sigma\sim\nu^3
\ee
where $\nu$ is a typical VEV of the fields. In the NMSSM wall solutions
may be found numerically\cite{aw}, and we find that the above equations
hold very well, with $\nu$ being replaced by some typical VEV of the
singlet which is of roughly the same magnitude as that of the Higgs
which gives the W and Z bosons their masses, {\em i.e.} 174GeV.

\subsection{Dynamics}
Once a wall network has formed, it will not remain static. There are
three important forces acting on the network of domain walls, namely
surface tension, friction, and pressure. The first of these is simply
the effect of the constant surface energy density, which makes it
energetically favourable for small bubbles of wall to disappear, for
walls to smooth themselves out, and so on. This removes the smaller
scale structure, and given enough time will remove the wall network
entirely. However, by causality it is clear that the largest possible
correlation length, by which we mean typical domain size, cannot be
larger than the horizon scale, and so the best we can expect is that
there is typically one wall per horizon volume throughout the evolution
of the universe. As will be discussed later, even if the entire visible
universe were to consist of only part of one domain, this still has
unwelcome cosmological consequences.

The second force is that of friction. As the wall moves through space
it will interact with the thermalised plasma, dissipating its energy
and slowing its evolution. This  effect can be shown to be
insignificant for very high temperatures, when the particles in the
plasma have very small reflection coeffecients and so do not exert a
large force, and also at low temperatures when the density of the
plasma is low\cite{asw}. We can thus neglect it, as it will only slow
down the wall evolution in its early stages.

A final force which we may consider is that of pressure. If one of the
vacua is slightly deeper than the others by some amount $\varepsilon$,
as a consequence of some slight {\em explicit} violation of the $\Z_3$
symmetry in the potential, then there will be a force per unit area on
the walls of order $\varepsilon$. Since the magnitude of the force due
to surface tension is given by $\sigma/R$, where $R$ is the
curvature scale (effectively the correlation length), we see that the
pressure will dominate the dynamics when
\be
\varepsilon>\frac{\sigma}{R}
\ee
Since $R$ is steadily increasing as a result of surface tension (and
the expansion of the universe), we see that for large enough values of
$\varepsilon$ the pressure will come to dominate the dynamics before
the present day. In this case the evolution of the wall network will
begin with the curvature scale increasing such that it is always of
order the horizon size, until ultimately pressure takes over, the true
vacuum comes to dominate, and the wall network is completely
eliminated. Numerical simulations suggest that once the pressure comes
to dominate, disappearance of the walls is almost as fast as causality
will allow\cite{sims}.

\subsection{Cosmological Implications}
We now turn to the cosmological implications of a domain wall network.
As the universe expands, the energy density due to the walls will fall
as $a^{-1}$ (for static walls) or $a^{-2}$ (for ultra-relativistic
walls in a radiation dominated universe), where $a$ is the cosmological
scale parameter. Since the energy densities of matter and radiation
fall as $a^{-3}$ and $a^{-4}$ respectively, it is clear that the walls
will ultimately dominate the dynamics of the universe, and we can check
that this will have happened long before the present day for weak scale
walls, and so the wall network must have long since disappeared.

In fact we may draw tighter constraints from other cosmological
observations. If the walls decay after nucleosynthesis, then they will
release a vast amount of energy which can be shown to cause
photodestruction of the light elements whose abundances can be
accurately measured\cite{nucbound}. Thus we require that the walls
decay before a temperature of around $0.1$ to 1MeV, giving a minimum
value of $\varepsilon$ to be
\be
\varepsilon\stackrel{>}{\sim}\lambda^{\prime}\sigma M_W^2/M_{Pl}
\ee
where $\lambda^{\prime}$ is of order $10^{-7}$. We conclude that a tiny
contribution to the superpotential from an NRO of form, say,
$\lambda^{\prime}N^4/M_{Pl}$ is sufficient to evade the cosmological
constraints. Such an NRO will have negligible effect on the low energy
phenomenology, and so we conclude that an NRO suppressed by at most one
power of the Planck mass is sufficient to eliminate the walls in time to
avoid cosmological problems.

\section{Destabilising Divergences}
The primary motivation for supersymmetry is the hierarchy problem. If
the standard model is an effective theory valid to a large scale
$\Lambda$ then we expect the hierarchy to be destabilised by quadratic
divergences and the standard model masses to be driven up to order
$\Lambda$. Supersymmetry prevents this happening, since softly broken
supersymmetry does not have quadratic divergences.

There is one important exception to this statement, in that tadpole
diagrams in supersymmetry may be quadratically divergent\cite{destab}
so long as we have NROs in our theory. These diagrams may be ruled out
for one of three reasons: gauge invariance, which cannot save us from
singlet tadpoles; $\Z_3$ symmetry, which our NROs must break to save us
from the cosmological implications; and supersymmetry which is
obviously broken for phenomenological reasons.

It is straightforward to check that every dimension five NRO which we
can introduce in our superpotential allows the construction of at least
one diagram at three loops or less which will be quadratically
divergent. For example, the least dangerous such operator is
$\lambda^{\prime}(H_1H_2)^2/M_{Pl}$, where the first such diagram
occurs at three loops. These diagrams generate a term in the
superpotential of form $\mu^{\prime\prime}N$, where
\be
\mu^{\prime\prime}\sim\frac{\lambda^{\prime}\lambda}{(16\pi^2)^2}
     m_{3/2}M_{Pl}
\ee
The presence of one factor of $m_{3/2}$ is guaranteed by the remark
above that in the absence on supersymmetry breaking the
non-renormalisation theorem prevents the generation of such terms,
while the factor $M_{Pl}$ is generated by taking the cut-off scale
$\Lambda\sim M_{Pl}$.

The effects of such terms on the hierarchy is catastrophic, since they
generate terms in the low energy Higgs potential which may be as large
as those from radiative breaking at each order, and so destroy any
reason for us to expect the electroweak scale to be very much less than
the GUT or Planck scales. Requiring that the electroweak scale is not
driven up by orders of magnitude in scale requires
$\lambda^{\prime}\stackrel{<}{\sim}10^{-11}$, in clear contrast to the
constraint from cosmology given above.

\section{Conclusions}
We have thus seen that the standard solution to the domain wall problem
in the NMSSM must always introduce destabilising divergences which will
destroy all predictive power in the Higgs sector and drive the
electroweak scale to be many orders of magnitude too large. One may
still solve the domain wall problem by introducing renormalisable terms
which break the $\Z_3$, such as a $\mu$ term, but one then has a far
more severe naturalness problem than is usually the case in the MSSM,
since we must somehow ban all the dangerous NROs which can cause
quadratically divergent singlet tadpoles, while simultaneously breaking
every symmetry which could ban them.

\end{document}